\documentclass[aps,pra,twocolumn,showpacs,amsmath]{revtex4-1}

\usepackage[per=slash,emulate=units,range-phrase = ...,range-units = single,exponent-product = \cdot,inter-unit-product = \cdot,separate-uncertainty = true]{siunitx}
\usepackage{hyperref}
\usepackage{graphicx}

\begin{document}

\title{A high resolution ion microscope for cold atoms}

\author{Markus Stecker}
\email[]{markus.stecker@uni-tuebingen.de}
\affiliation{Center for Quantum Science, Physikalisches Institut, Eberhard Karls Universit\"at T\"ubingen, Auf der Morgenstelle 14, D-72076 T\"ubingen, Germany}
\author{Hannah Schefzyk}
\affiliation{Center for Quantum Science, Physikalisches Institut, Eberhard Karls Universit\"at T\"ubingen, Auf der Morgenstelle 14, D-72076 T\"ubingen, Germany}
\author{J\'{o}zsef Fort\'{a}gh}
\affiliation{Center for Quantum Science, Physikalisches Institut, Eberhard Karls Universit\"at T\"ubingen, Auf der Morgenstelle 14, D-72076 T\"ubingen, Germany}
\author{Andreas G\"unther}
\email[]{a.guenther@uni-tuebingen.de}
\affiliation{Center for Quantum Science, Physikalisches Institut, Eberhard Karls Universit\"at T\"ubingen, Auf der Morgenstelle 14, D-72076 T\"ubingen, Germany}

\date{\today}

\begin{abstract}
We report on an ion-optical system that serves as a microscope for ultracold ground state and Rydberg atoms. The system is designed to achieve a magnification of up to 1000 and a spatial resolution in the \SI{100}{\nm} range, thereby surpassing many standard imaging techniques for cold atoms. The microscope consists of four electrostatic lenses and a microchannel plate in conjunction with a delay line detector in order to achieve single particle sensitivity with high temporal and spatial resolution. We describe the design process of the microscope including ion-optical simulations of the imaging system and characterize aberrations and the resolution limit. Furthermore, we present the experimental realization of the microscope in a cold atom setup and investigate its performance by patterned ionization with a structure size down to \SI{2.7}{\micro\meter}. The microscope meets the requirements for studying various many-body effects, ranging from correlations in cold quantum gases up to Rydberg molecule formation.
\end{abstract}


\maketitle

\section{Introduction}\label{sec:intro}

The development of efficient laser cooling \cite{Chu.1985} paved the way to the production of ultracold atom clouds with temperatures down to several nanokelvin and finally to the experimental realization of Bose-Einstein condensates \cite{Anderson.1995, Davis.1995} and degenerate Fermi gases \cite{DeMarco.1999}. In order to study the unique properties of ultracold quantum matter, various techniques have been established. 

The standard way to image ultracold atom clouds is absorption imaging \cite{Ketterle.1999}, where the cloud is illuminated by a resonant laser beam and the shadow image is recorded on a camera. However, the cloud is destroyed by this method. The destruction can be suppressed when using the related method of phase contrast imaging, allowing to image the same cloud several times \cite{Andrews.1996,Higbie.2005,Meppelink.2010}, yet at a reduced contrast.

For non-destructive imaging of dilute samples with single atom sensitivity, fluorescence imaging is typically used. Here, non-destructive means that the atoms are not lost out of the trap during the imaging process. In general, this type of imaging requires relative long exposure times up to \SI{1}{s} to gather enough fluorescence photons. During this time the atoms may heat up, such that single atom resolution is typically achieved in deep trapping structures. With cold atoms loaded into a two dimensional optical lattice and a light-optical system with high numerical aperture placed close to the atom sample, a spatial resolution below \SI{1}{\micro\meter} could be achieved with single atom and single site sensitivity \cite{Bakr.2009,Sherson.2010}. These ``quantum gas microscopes" revolutionized the capabilities of detection and control of quantum many body systems and allow for example the direct observation of the superfluid to Mott insulator phase transition on the microscopic level \cite{Bakr.2010,Sherson.2010}. Recently, this technique was extended from bosonic to fermionic quantum matter \cite{Edge2015,Cheuk2015,Haller.2015,Greif.2015},
thus providing an ideal quantum simulator for investigating solid-state Fermi systems that are difficult to analyze numerically \cite{Troyer.2005,Parsons.2016}. These microscopes promise important insights into open questions such as high-temperature superconductivity.

Above methods work well for closed cycle transitions, where multiple photons within the visible or near infrared regime can be scattered. If not ground state atoms but Rydberg atoms (atoms in states of high principal quantum number \cite{Gallagher.1994}) shall be imaged, the imaging techniques cannot always be applied directly. In case of earth alkali Rydberg atoms, the second valence electron can be used for such a closed cycle \cite{McQuillen.2013}. But with the more commonly used alkali Rydberg atoms, only an indirect approach is possible, by detecting losses in the ground state population \cite{Urban.2009, Gaetan.2009} or by transferring the Rydberg population to the detectable ground state \cite{Schauss.2012}. Thus, alkali Rydbergs are preferably detected by ionization \cite{Gallagher.1994} - which also works for ground state atoms \cite{Stibor.2010} - or alternatively by the all-optical approach of electromagnetic induced transparency \cite{Gunter.2012, Karlewski.2015}.

The detection of ions or electrons with high temporal and spatial resolution can be achieved with microchannel plates (MCP). Such detectors have been used to image cold electron or ion beams extracted out of a laser-cooled atom cloud \cite{Claessens.2007,Hanssen.2008,Reijnders.2009,McCulloch.2011,McCulloch.2013}. Similar to a field ion microscope, spatially resolved detection of Rydberg atoms has been demonstrated by ionization and acceleration in the divergent electrical field of a metallic nanotip with a magnification of 300 and a resolution better than $\SI{1}{\micro\meter}$ \cite{Schwarzkopf.2011, Schwarzkopf.2013}. Imaging of spatially structured photoionization and Rydberg excitation patterns in a magneto-optical trap could be achieved with a magnification of 46 and a resolution of 10-\SI{20}{\micro\meter} \cite{Bijnen.2015}. In a different approach, a spatial resolution of \SI{100}{nm} has been realized, using a scanning electron microscope for local ionization of cold atoms \cite{Gericke.2008, Manthey.2014}. All these methods image a two-dimensional plane. However, a MCP can also be used for three-dimensional detection, with the third dimension being calculated out of the timing information, as demonstrated with a Bose-Einstein condensate of metastable helium \cite{Schellekens.2005}.

Here, we present a novel ion microscope for in-situ and real-time imaging of ultracold atomic gases. The microscope features a full ion-optical imaging system with four electrostatic lenses and a MCP of \SI{40}{mm} diameter for ion detection. Using the MCP in conjunction with a delay line detector, single particle sensitivity with high temporal and spatial resolution is achieved. Depending on the ionization scheme, the system may be used for detecting ground state atoms as well as Rydberg atoms. The ion microscope has been optimized for detecting cold atomic ensembles with typical extensions in the 10 to \SI{100}{\micro\meter} range. Therefore, the magnification can be adjusted between 10 and 1000 with a theoretical resolution limit of \SI{100}{nm}. Ion-optical abberations have been minimized to provide a distortion-free imaging over the whole field of view. A large depth of field ensures the possibility to reckon back from the timing information onto the third spatial direction. We demonstrate the microscope using a continuously operated magneto-optical trap (MOT) of rubidium atoms. In order to test the performance of the microscope, different optical ionization patterns are imprinted onto the MOT, with the generated ions being imaged by the electrostatic lens system.

\section{Ion optics}\label{sec:sim}

In analogy to light optics, where light beams can be refracted by materials with different refractive indices, the trajectories of charged particles can be manipulated by electromagnetic fields. The trajectories of ions are - compared to electrons - less sensitive to magnetic fields, so we use electrostatic fields for our ion-optical system. Our setup consists of a set of four einzel lenses - a type of lens that is typically used for low-energy ions \cite{Drummond.1984}. Each einzel lens consists of three consecutive, rotational symmetric electrodes with an aperture. Usually, the two outer electrodes are held at the same electric potential, so the energy of the charged particle after exiting the lens is equal to its energy before entering. By changing the potential of the inner electrode, the focal length of the lens can be varied.

For the design of the ion optics, we made use of a commercial field and particle trajectory simulation program. The program calculates the electrostatic potential for a given electrode geometry by applying a finite difference method and then simulates charged particle trajectories in this potential. All measures like aperture sizes, lens separations and electrode lengths have been optimized to meet the microscope's target specifications with a maximum magnification of 1000 and a resolution better than \SI{100}{nm} (cf. section \ref{sec:sim_res}).

\begin{figure*}
	\includegraphics{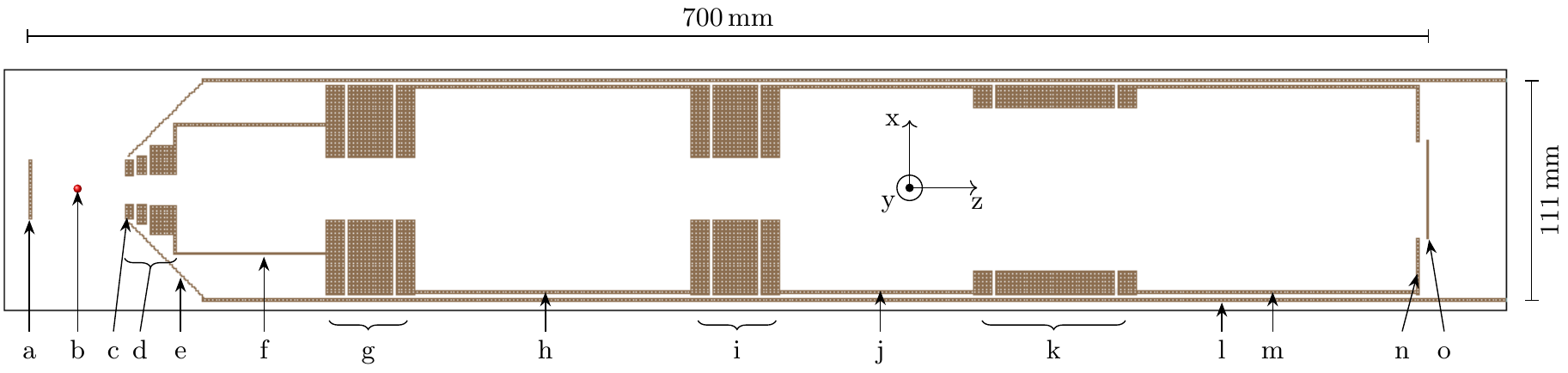}
	\caption{\label{fig:setup_ionenoptics_schematic} Cross section (to scale) of the rotational symmetric electrode geometry: a) positive extraction electrode, b) center of MOT, c) negative extraction electrode, d) einzel lens L1 (consisting of three electrodes including the negative extraction electrode), e) grounded shielding cone, f) drift tube 1, g) einzel lens L2, h) drift tube 2, i) einzel lens L3, j) drift tube 3, k) einzel lens L4, l) grounded shielding tube, m) drift tube 4, n) MCP shielding, o) MCP stack. }
\end{figure*}

The microscope is realized in a \SI{700}{mm} long tube with a diameter of \SI{111}{mm} (see Fig. \ref{fig:setup_ionenoptics_schematic} and Tab. \ref{tabular1}).
\begin{table*}
\begin{center}
 \begin{tabular}{ | l | l | l | l |l |}
    \hline
    \textbf{element}&\textbf{aperture radius} & \textbf{outer radius} & \textbf{length} &\textbf{distance} \\ \hline
    positive extraction electrode (a) & -  & \SI{15}{mm} & \SI{2}{mm} & -\\ \hline
    
    negative extraction electrode (c) & \SI{6.5}{mm}  & \SI{15}{mm} & \SI{5}{mm} & \SI{46}{mm}\\ \hline
    center electrode of L1 (d) & \SI{7}{mm}  & \SI{17.5}{mm} & \SI{5}{mm} & \SI{1}{mm} \\ \hline
    rear electrode of L1 (d)  & \SI{7.5}{mm} & \SI{22.5}{mm} & \SI{14.5}{mm} & \SI{1}{mm}\\ \hline
    
    drift tube 1 (f) & \SI{31}{mm}  & \SI{32}{mm} &  \SI{74}{mm} & \SI{0}{mm} \\ \hline
    
    front electrode of L2 (g)   & \SI{15}{mm} & \SI{52.5}{mm} & \SI{10}{mm} & \SI{0}{mm}\\ \hline
    center electrode of L2 (g) & \SI{15}{mm}  & \SI{49}{mm} & \SI{22.5}{mm} & \SI{1}{mm}\\ \hline
	rear electrode of L2 (g) & \SI{15}{mm} & \SI{52.5}{mm} & \SI{10}{mm}& \SI{1}{mm}\\ \hline
    
    drift tube 2 (h) & \SI{50}{mm}  & \SI{51.5}{mm} & \SI{137.5}{mm} & \SI{0}{mm}\\ \hline
    
    front electrode of L3 (i)  & \SI{15}{mm} & \SI{52.5}{mm} & \SI{10}{mm}  & \SI{0}{mm}\\ \hline
	center electrode of L3 (i) & \SI{15}{mm}   & \SI{49}{mm}  & \SI{22.5}{mm} & \SI{1}{mm}\\ \hline
	rear electrode of L3 (i)  & \SI{15}{mm} & \SI{52.5}{mm}& \SI{10}{mm}& \SI{1}{mm}\\ \hline
    
   	drift tube 3 (j) & \SI{50}{mm} & \SI{51.5}{mm} & \SI{96.5}{mm} & \SI{0}{mm} \\ \hline
    
    front electrode of L4 (k)  & \SI{40}{mm} & \SI{52.5}{mm} & \SI{10}{mm} & \SI{0}{mm}\\ \hline
    center electrode of L4 (k) & \SI{40}{mm} & \SI{49}{mm} & \SI{60}{mm} & \SI{1}{mm}\\ \hline
	rear electrode of L4 (k)   & \SI{40}{mm} & \SI{52,5}{mm} & \SI{10}{mm} & \SI{1}{mm}\\ \hline
    
	drift tube 4 (m) & \SI{50}{mm} & \SI{51.5}{mm} & \SI{138.5}{mm} & \SI{0}{mm}\\ \hline
 
	MCP shielding (n) &  \SI{23.5}{mm}&  \SI{51.5}{mm}& \SI{2}{mm} & \SI{0}{mm}\\ \hline
	MCP stack (o) & - & \SI{25}{mm} (\SI{20}{mm} active) & \SI{3}{mm} & \SI{3}{mm} \\ \hline
   
    \hline
    \end{tabular}
\end{center}
\caption{Dimensions of the ion-optical elements. The given distances are meant as clear distance to the preceding element. Front electrodes correspond to electrodes facing towards the MOT, rear electrodes towards the MCP. The labels (a)-(o) refer to Fig. \ref{fig:setup_ionenoptics_schematic}.}
\label{tabular1}
\end{table*}
The object, here an atom cloud in a MOT, is centered in between a pair of extraction electrodes. These are typically held at $U_{\text{ext}} = \SI{\pm 500}{\volt}$ and form the electric field to extract the ions generated in the MOT. Four consecutive einzel lenses image the ions onto a microchannel plate detector with \SI{40}{\mm} diameter. Starting with einzel lens 1, each lens produces a magnified image in the image plane behind the lens, which is further magnified by the next lens and finally imaged onto the MCP plane. In order to achieve a sharp image, the focal length of each lens is matched to the position of the image plane of the previous lens. The position of the image planes is determined by simulating different ion trajectories coming from a common starting point and determining the intersection of the trajectories behind the lens. There are of course several voltage settings for the different lenses to achieve a sharp image for a specific magnification but typically the refractive power should be distributed homogeneously over the lenses to reduce aberrations.

The drift tubes between the lenses and the outer electrodes of the lenses 2, 3 and 4 are held at \SI{-2.4}{\kilo\volt} which equals the potential at the front plate of the MCP detector. The drift tubes keep the lenses at a fixed distance and ensure an undisturbed (field free) particle movement in between the lenses. In contrast to the other lenses, einzel lens 1, with the negative extractor serving as its first electrode, acts as an immersion lens, accelerating the ions to the drift tube potential. The voltage at the inner electrodes of the einzel lenses can be adjusted to change the focal length of the lens and with that the magnification of the system. In general, a higher voltage leads to a higher magnification. The inner electrode of the first einzel lens is typically held at $U_{\text{L1}} = \SIrange{-2.7}{-3}{\kilo\volt}$, the inner electrodes of the three other lenses at $U_{\text{L2,3,4}} = \SIrange{0}{+800}{\volt}$.

\begin{figure}
	\includegraphics{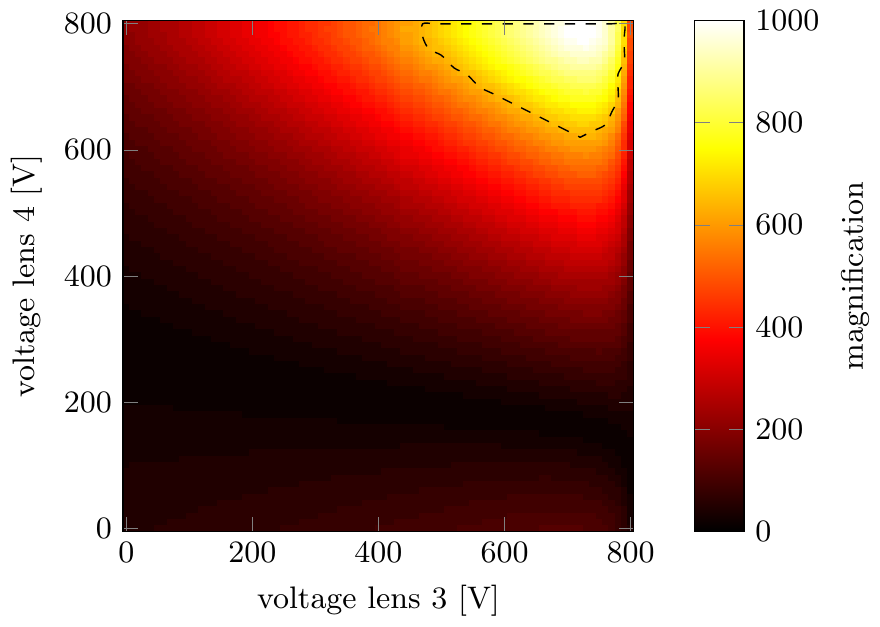}
	\caption{\label{fig:mag_vs_U3U4}Magnification of the ion optics at the MCP plane for different voltages at the inner electrodes of einzel lens 3 and 4. The remaining parameters were fixed to $U_{\text{ext}}=\pm \SI{500}{V}$, $U_{\text{L1}}=\SI{-2700}{V}$ and $U_{\text{L2}}=\SI{+750}{V}$. All data is derived from a numerical simulation. The sharpest images (smallest point spread function) can be achieved with voltage combinations inside the dashed line. For a sharp image at other magnifications, the voltages at the lenses L1 and L2 have to be adjusted accordingly.}
\end{figure}

To determine the achievable magnification, we simulated the imaging of ion patterns positioned in the center between the extractor electrodes onto the MCP plane. In Fig. \ref{fig:mag_vs_U3U4}, the magnification of the system depending on the voltages at einzel lens 3 and 4 is shown.
The simulations predict the desired magnification of 1000 for a voltage at the einzel lenses 2, 3 and 4 of $U_{\text{L2,3,4}}\simeq\SI{750}{V}$ (with $U_{\text{ext}}=\pm \SI{500}{V}$ and $U_{\text{L1}}=\SI{-2700}{V}$). The time of flight of the ions also varies with electrode voltages and ranges in between 10 and \SI{14}{\micro\second}.

\section{Aberrations and resolution limit}
\label{sec:sim_res}
The resolution of an ion-optical system is, in analogy to light optical systems, in general limited by aberrations. In the following section, the types of aberrations that have a relevant influence on our imaging system are covered. Independent of the properties of the ion-optical imaging system, the resolution is limited by the MCP ion detector itself, which has a center-to-center pore distance of \SI{17}{\micro\meter}. Therefore, following the Nyquist-Shannon sampling theorem \cite{Shannon.1949}, only structures of size $\geq 2 \times \SI{17}{\micro\meter}$ can be resolved. Thus, every aberration effect has to be compared to this principal limit.

\subsection{Spherical aberrations}
Spherical aberrations in rotational symmetric lens systems are unavoidable \cite{Scherzer.1936}, but can be limited by splitting the total refractive power on several lenses, as realized in our microscope. This reduces the angle of incidence of the ion trajectories at each lens and minimizes spherical aberrations. Besides this, thorough optimization of the lens parameters is required to reduce spherical aberrations further.
 
With the outer lens electrodes being at the same potential as the drift tubes, their length does not play a crucial role for the imaging system. Basically, the refraction takes place in the interspaces between the outer electrodes and the central electrode, where the potential landscape is changing rapidly. The separation between the electrodes has been fixed to \SI{1}{mm} to ensure a sufficient dielectric strength (see sec. \ref{sec:real}). This leaves the applied voltages, the aperture sizes, and the lengths of the central electrodes for optimization. They all have direct impact on the refractive power of the lens and thus on the focal length and magnification. On the other hand, they strongly influence the spherical abberations.
 
With the spherical abberations being inverse proportional to the aperture sizes \cite{Szilagyi.1986}, large apertures are generally preferable. For large apertures, however, the focal length and the principal distance of the image plane are increased. This can be partially compensated by increasing the applied voltage to the central electrode. Aperture sizes are thus limited by the target length of the microscope and the maximal applicable voltages. For our imaging system, we limited the size to \SI{700}{mm} and the voltage difference between neighboring electrodes to \SI{3.2}{kV}.
 
The aperture sizes cannot be optimized independently of the lengths of the central electrodes. For increasing electrode length, the focal length and the magnification is reduced. As with the aperture size, the central electrode length has direct influence onto the spherical abberations. They can be minimized by minimizing the gradient of the electrostatic potential along the optical axis at the entrance and exit of the lens. This means that the electrode has to be long compared to the aperture \cite{Sise.2005}. In our setup, the optimization resulted in a length of the central electrode being typically 50\% larger than its aperture radius.
 
\begin{figure*}
	\includegraphics{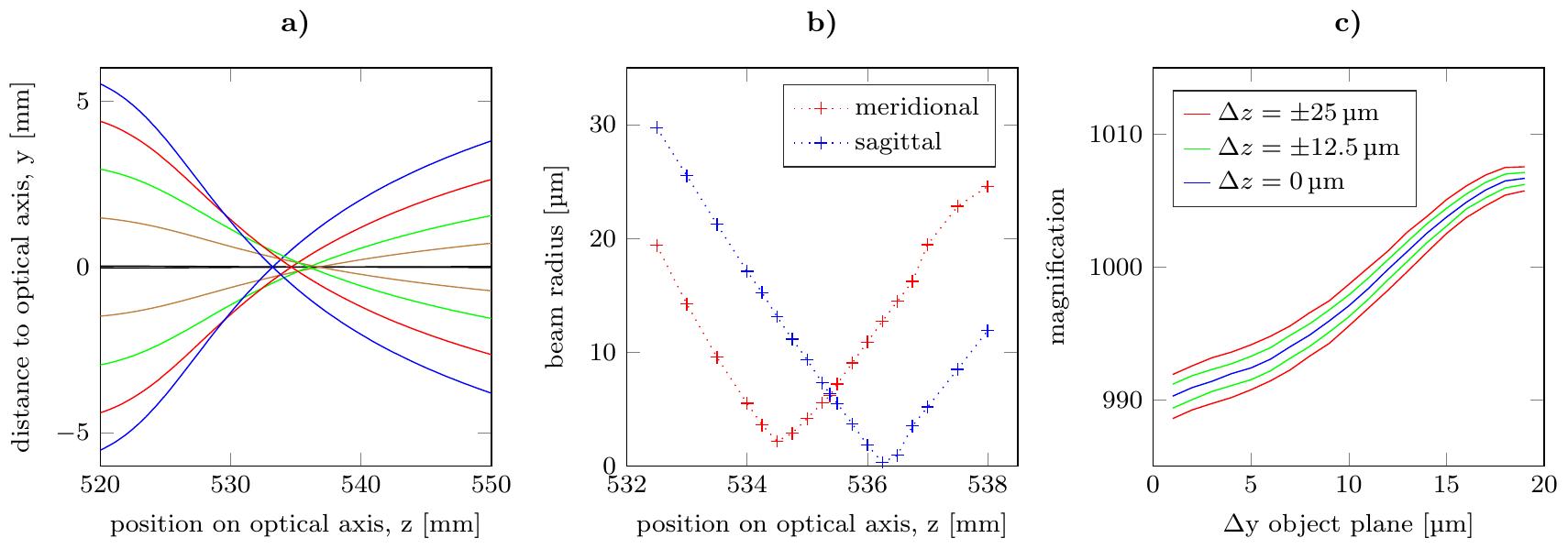}
	\caption{\label{fig:sim_spherical_astigm} Monochromatic abberations of the ion-optical system at a magnification of 1000. a) Spherical aberration: ion trajectories near the focus of einzel lens 4. The different lines correspond to different starting distances of the ion to the optical axis (from outer to inner lines: $\Delta y=\pm\SI{19}{\micro\meter},\pm\SI{15}{\micro\meter},\pm\SI{10}{\micro\meter},\pm\SI{5}{\micro\meter},\pm\SI{100}{\nano\meter}$). These result in different focal points and a broadened focus size. b) Astigmatism: radius of an ion beam starting with a diameter of \SI{100}{\nano\meter} and a distance of \SI{10}{\micro\meter} to the optical axis in meridional and sagittal direction. Depicted is the beam radius after refraction by einzel lens 4. c) Distortion: magnification as a function of the object distance to the optical axis. Different colors correspond to different object positions along the optical axis.}
\end{figure*}

The remaining effects of spherical aberration in our system can be seen in Fig. \ref{fig:sim_spherical_astigm}a.
It shows the focus region of einzel lens 4 for ion trajectories with different starting points. The focal length for off-axis beams is considerably shorter than for beams close to the optical axis.

\subsection{Astigmatism and distortion}
The effects of astigmatism and distortion are also an inherent feature of einzel lenses \cite{Klemperer.1953}.
 
Astigmatism means, that the meridional section (the plane containing the object point and the optical axis) of a beam of rays far away from the optical axis has a different focal length than the sagittal section (the plane perpendicular to the meridional plane). This can lead to a different imaging resolution of the ion optics in sagittal and meridional direction. In Fig. \ref{fig:sim_spherical_astigm}b, the effect of astigmatism on our ion-optical imaging is visualized: the position of the focal point differs between meridional and sagittal section of the beam. Also the beam radius at the focal point and therefore the resolution of the imaging system is different for the two directions. This effect of directional resolution can also be seen in the simulated imaging of ion test structures in Fig. \ref{fig:simulated_teststructures}.
 
Distortion means, that the magnification of the optics changes with increasing distance from the optical axis. Its influence is shown in Fig. \ref{fig:sim_spherical_astigm}c for different positions of the object plane with respect to the central position between the two extraction electrodes. In principal, the magnification changes by about $1.5\%$ across the image plane, resulting in a superposition of barrel and pincushion distortion.

\subsection{Chromatic aberrations}
\label{sec:sim_chrom}
Besides these monochromatic aberrations, there are also chromatic aberrations effecting the imaging quality. In general, the focal length decreases for ions with lower energy, as the refractive power of the lens is increased. This makes the starting energy distribution of the ions to be one of the main sources for chromatic aberrations. For the experiments described here, the ions are produced via photoionization of rubidium atoms out of a continuously operated magneto-optical trap (MOT) (c.f. sec. \ref{sec:real}). The starting energy distribution is thus given by the temperature of the MOT, the excess energy (the difference between the photon energy and the ionization energy of rubidium) and the momentum transfer from the photoionization process. The MOT temperature is typically around $T=\SI{100}{\micro\kelvin}$ which corresponds to a particle energy of $E_{\text{kin}}=k_BT=\SI{8.6}{neV}$. Following energy conservation, the excess energy is distributed on the electron and the ion correspondingly to their mass ratio. With $m_{Rb^+}/m_e \approx 1.6\times 10^5$ less than 0.001\% of the energy goes to the ion. With the ionization laser being tuned close to the ionization threshold, the excess energy can be very small, in our case $\SI{104}{\micro eV}$, which leads to an energy transfer onto the ion of $\SI{0.65}{neV}$. The contribution of the momentum transfer from the photon can be neglected as it is more than one order of magnitude smaller than the other two. This leaves the thermal energy of the particles and the excess energy to be the dominant contribution to the chromatic aberration. Our ion-optical system is designed such that starting energies up to \SI{2.5}{\micro \electronvolt} will not affect the minimal resolution. This ensures optimal imaging conditions for cloud temperatures up to \SI{30}{mK} or total excess energies up to \SI{0.4}{eV}.

\subsection{Depth of field}
\begin{figure*}
	\includegraphics{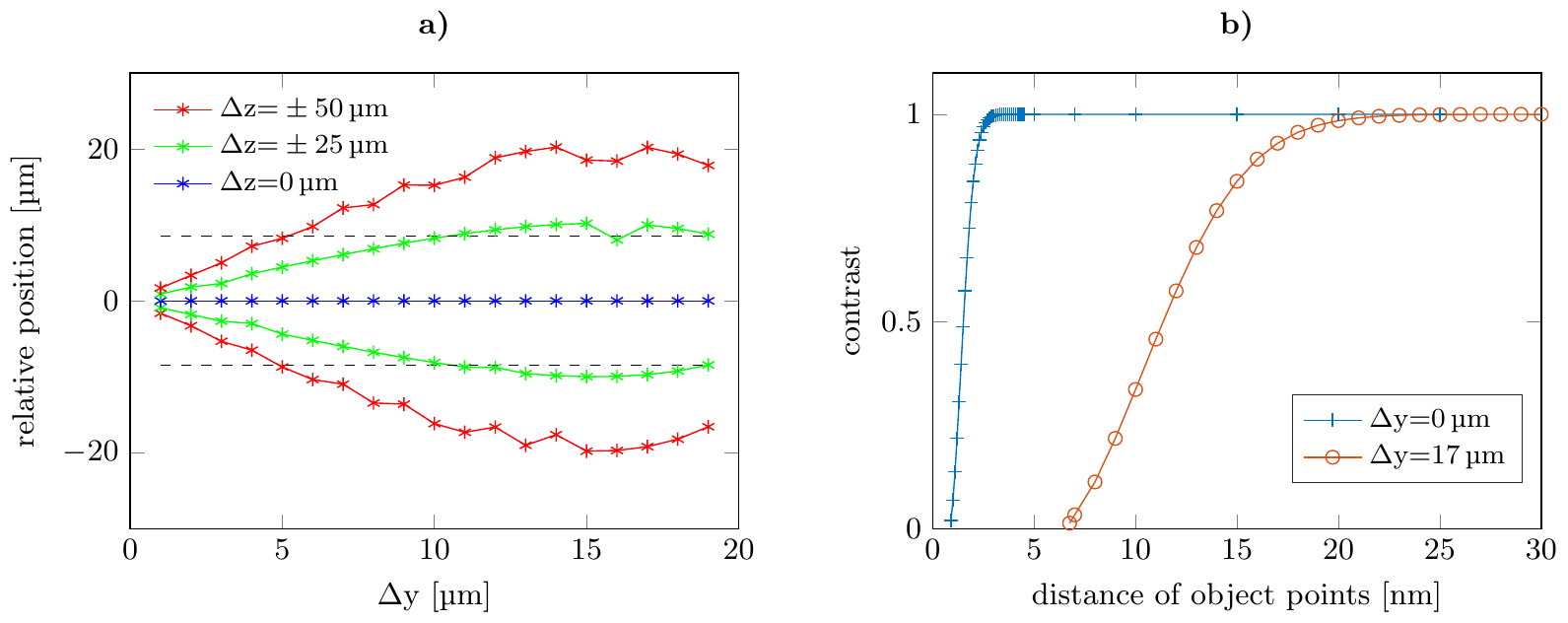}
	\caption{\label{fig:depth_of_field} a) Depth of field: radial position of the ion trajectories at the detector plane for various starting positions along the optical axis $\Delta z$ and distances to the optical axis $\Delta y$ in the object plane. $\Delta y=\Delta z=0$ is the center point between the extraction electrodes. The position is shown relative to the point of incidence of an ion starting at the given $\Delta y$ but with $\Delta z=0$ (blue line). The dashed lines depict the MCP pore distance. b) Modulation transfer function: contrast in the image plane as a function of the point separation in the object plane, for on-axis ($\Delta y = 0$) and off-axis ($\Delta y = \SI{17}{\micro\meter}$) objects. The magnification is set to 1000.}
\end{figure*}

In addition to the chromatic aberration resulting out of the starting energy of the ions, there is also another source of varying ion energy: The ions are not only produced in a plane but in a three dimensional volume. Ions with different starting points along the optical axis experience different accelerations before they reach the first lens. The simulations show, that at a magnification of 1000 the spread of starting positions along the optical axis of the ions can be as big as \SI{50}{\micro\meter} without limiting the resolution to a value worse than the principal limit of the ion detector. At the same time, this value defines the depth of field for our imaging system. The effects are visualized in Fig. \ref{fig:depth_of_field}a, showing the relative position shift of ions in the detector plane due to varying starting positions along the optical axis.
For starting position shifts $\left|\Delta z\right| \leq \SI{25}{\micro\meter}$ the relative shift in the image plane stays within the pixel size of the MCP detector.

\subsection{Modulation transfer function and resolution} 
In summary, all of the above aberrations limit the final resolution of the ion-optical imaging system. This is typically quantified via the modulation transfer function, describing the contrast in the image plane as function of the point separation in the object plane. Fig. \ref{fig:depth_of_field}b shows the resulting modulation transfer function for our ion-optical imaging system with the magnification set to 1000 and neglecting the discretization due to the finite MCP pore size. The object depth along the optical axis has been chosen to be \SI{25}{\micro\meter}. For on-axis beams, full contrast can be transferred for point separations down to several nanometer. However, off-axis beams are limited to point separations in the \SI{10}{nm} regime.

\begin{figure*}
	\includegraphics{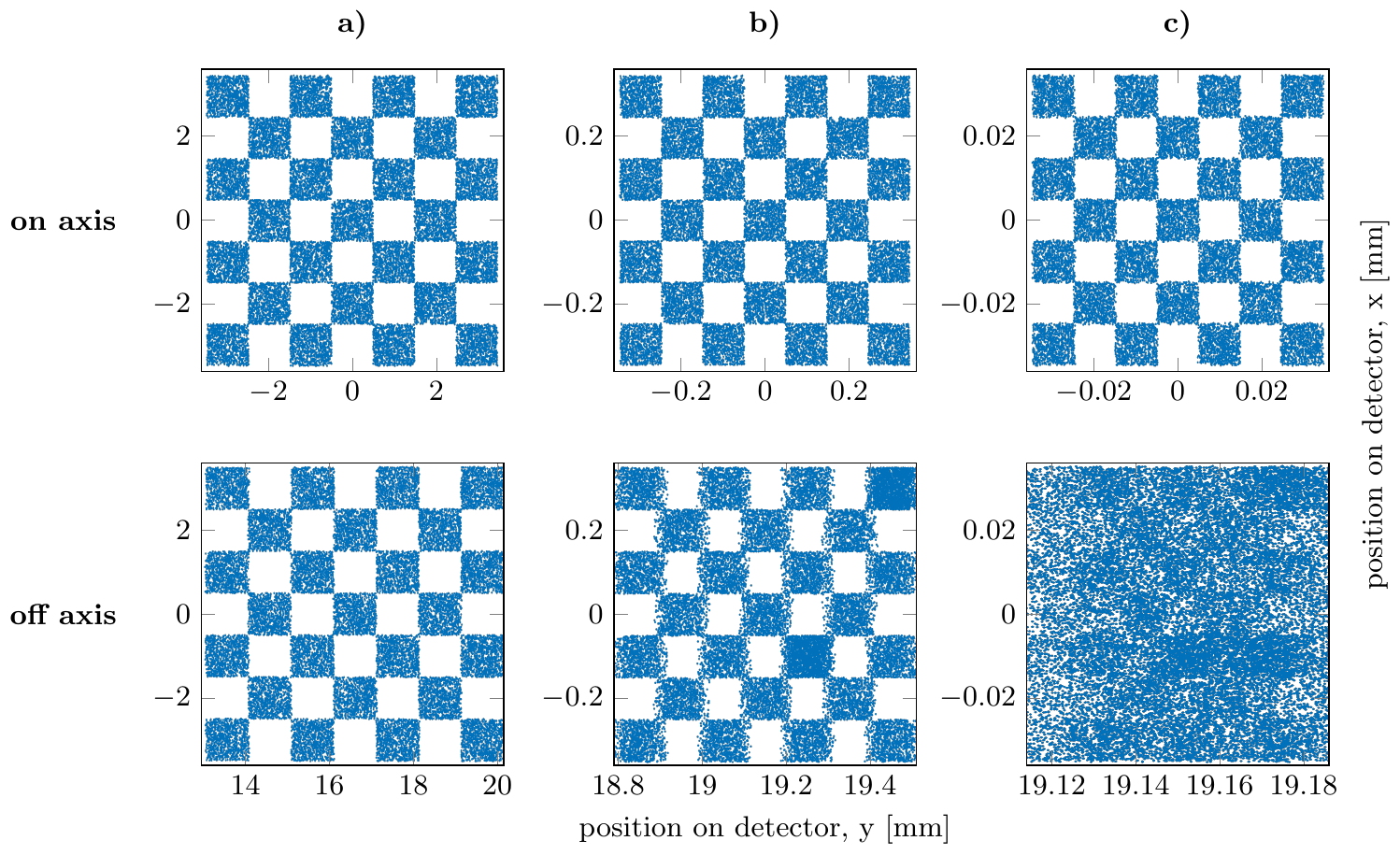}	
	\caption{\label{fig:simulated_teststructures} Simulated imaging of several test patterns near (top row) and \SI{19}{\micro\meter} away (bottom row) from the optical axis (ion optics set to a magnification of 1000). The test structures consist of cuboids of different sizes filled with ions: a) $\SI{1}{\micro\meter}\times\SI{1}{\micro\meter}$, b) $\SI{100}{\nano\meter}\times\SI{100}{\nano\meter}$, c) $\SI{10}{\nano\meter}\times\SI{10}{\nano\meter}$. The extent along the optical axis is always \SI{50}{\micro\meter}, the starting energy \SI{9.3}{neV}.}
\end{figure*}

The results can be nicely visualized by simulating the imaging of a test pattern (see Fig. \ref{fig:simulated_teststructures}).
The starting points of the simulated ions are randomly distributed in cuboids of different sizes placed in the center plane between the extraction electrodes. The extension of such a cuboid in the object plane (perpendicular to the optical axis) is small (down to \SI{10}{nm}) in order to visualize the resolution limit of the ion optics. In contrast to that, the extent along the optical axis is large (\SI{50}{\micro\meter}) to show the large depth of field of the system. The starting kinetic energy of each ion is \SI{9.3}{neV} (see section \ref{sec:sim_chrom}) with a random direction of the starting velocity. The trajectories of all ions are simulated and in the ideal case, the spatial distribution should be reproduced on the detector plane with the corresponding magnification. The results in  Fig. \ref{fig:simulated_teststructures} show that for structures close to the optical axis the imaging is nearly perfect, but further away from the optical axis, aberrations play a significant role in the imaging quality. With an object depth of $\Delta z = \SI{50}{\micro\meter}$ the \SI{100}{nm} structures can be nicely resolved, however, the \SI{10}{nm} structures are fully washed out for off-axis patterns. Here, the resolution limit is at about \SI{50}{nm}.

\subsection{Three-dimensional imaging} 
As the time of flight of the ion varies with its starting position, it is possible to calculate back from the timing information of the MCP on the initial z-position of the ion. This position along the optical axis is not directly accessible by the imaging. For the procedure to work, the ion has to be inside the depth of field of the imaging system because the time of flight also varies with the starting position in the object plane (x and y-direction). For the system set to a magnification of 1000, the time of flight increases quadratically with the distance from the image center. It is \SI{200}{ns} higher at the edge of the image than in the center. If this effect has been taken into account, the residual variation of the time of flight originates from the starting position along the optical axis. This variation is linear and has a gradient between \SI{0.244}{\micro\second\per\milli\meter} and \SI{0.315}{\micro\second\per\milli\meter} depending on the x/y-position of the ion. Therefore, the accuracy of the time of flight measurement has a direct influence on the spatial resolution in z-direction. With a timing resolution of the MCPs time to digital converter of $<\SI{100}{ps}$, the spatial resolution in z-direction is about \SI{400}{\nano\meter}. This is about an order of magnitude higher than the design resolution in x/y-direction. Similar results may be achieved using a movable light sheet for localizing the ionization process along the z-axis.

\section{\label{sec:real}Experimental realization}
\begin{figure*}
	\includegraphics{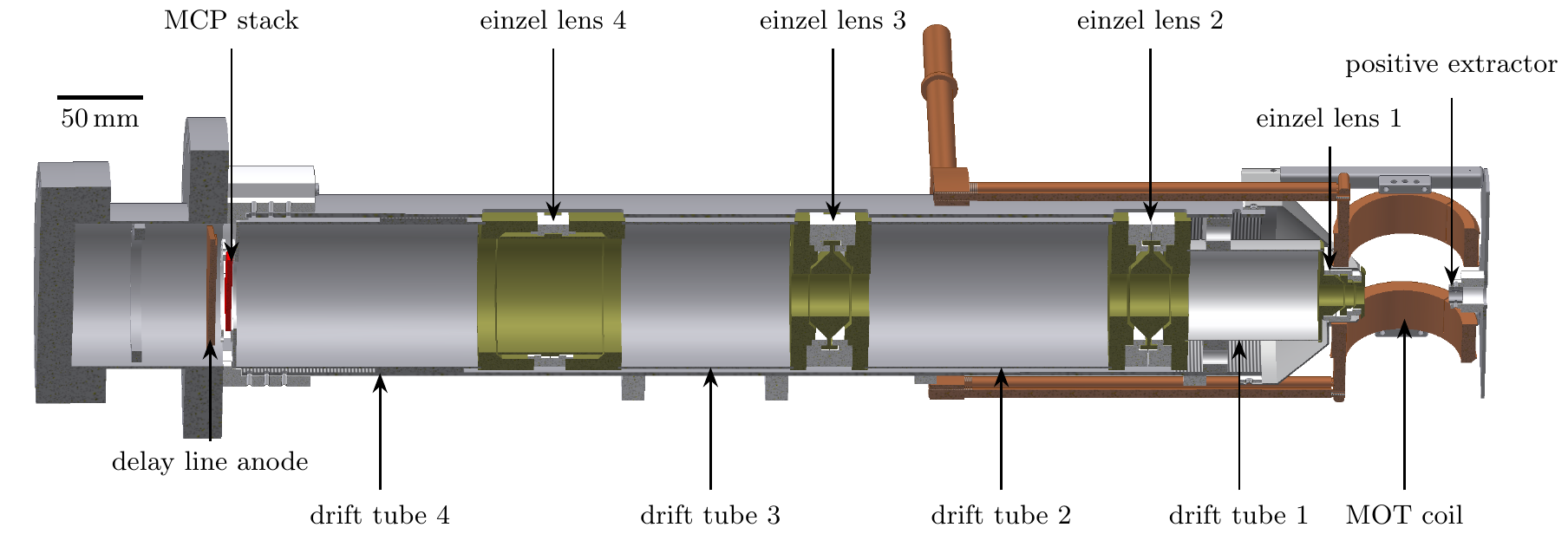}
	\caption{\label{fig:complete_setup}Cross-section of the full ion optics setup inside the vacuum chamber.}
\end{figure*}

The complete experimental setup is illustrated in Fig. \ref{fig:complete_setup}.
It can be divided into three parts: MOT, ion optics and ion detection. All parts are placed in a vacuum chamber under ultra-high vacuum conditions at a pressure of $\sim \SI{1e-10}{mbar}$.

\begin{figure}
	\includegraphics{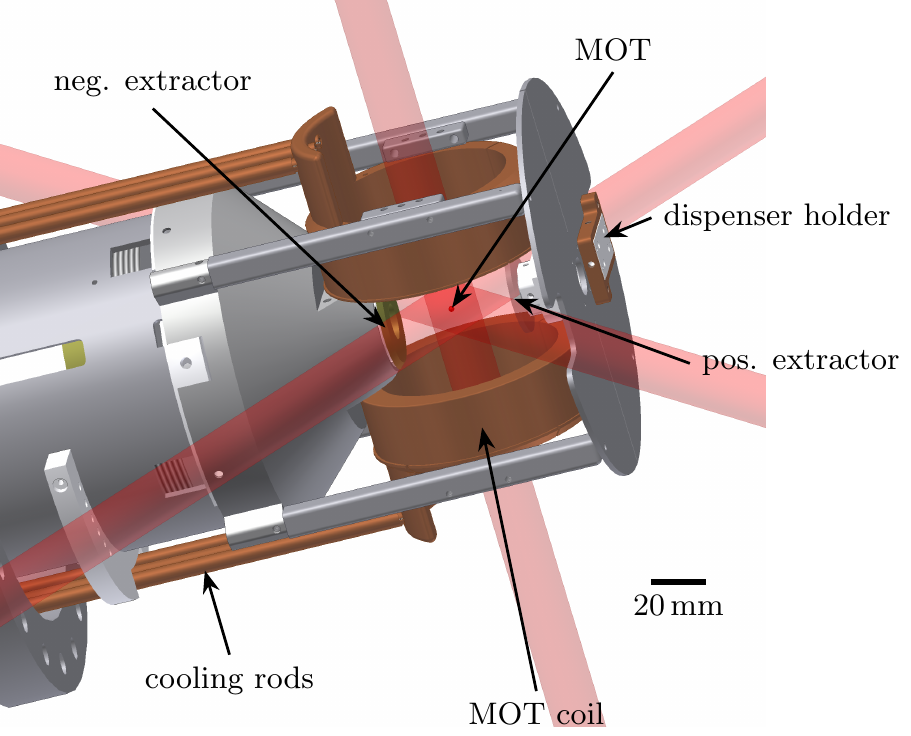}
	\caption{\label{fig:MOT_setup} Part of the setup where the magneto-optical trap (MOT) is formed by six intersecting laser beams and a magnetic field formed by two coils.}
\end{figure}

The MOT-part (see Fig. \ref{fig:MOT_setup}) is located between the two extractor electrodes and allows for optical access from several directions.
The MOT is created at the minimum of a magnetic quadrupole field generated by a pair of coils in anti-Helmholtz configuration. The magnetic field gradient along the coil axis is \SI{11}{G/cm}.  The radius of the coils (inside/outside: 34/\SI{42}{mm}) was chosen to be as big as possible to minimize the influence of the coils and their copper holders on the extraction field, while keeping the number of windings and current required for obtaining the field gradient to reasonable values. The laser cooling is provided by three pairs of counterpropagating laser beams which intersect at the magnetic field minimum. The MOT is loaded with rubidium 87 atoms that are provided by current-heated dispensers. 

Before imaging the cold atoms, they have to be ionized. This is done by a two-step  photoionization process. Using the cooling laser of the MOT, rubidium 87 ground state atoms are first excited from the $5S_{1/2}, F=2$ to the $5P_{3/2}, F=3$ state. Ionization is then achieved via a laser with wavelength of $\SI{479.04}{nm}$. The produced $\text{Rb}^+$ ions are then imaged with the ion optics.

\begin{figure}
	\includegraphics{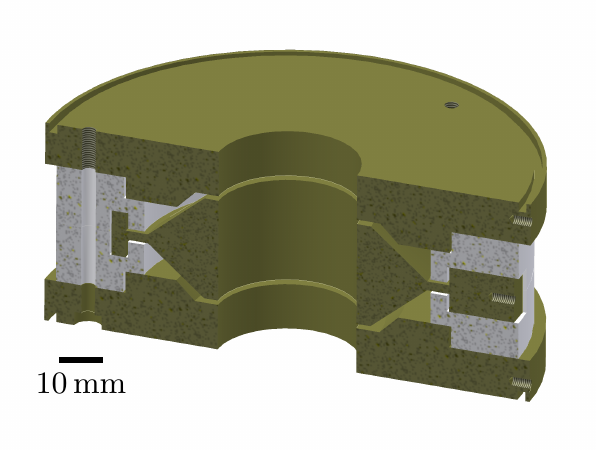}
	\caption{\label{fig:lens_crosssection} Cross section of einzel lens 3 with stainless steel electrodes (dark) and Macor insulators (light). The distance between two electrode surfaces is \SI{1}{mm} in vacuum and \SI{7.25}{mm} at  Macor insulators.}
\end{figure}

The ion optics was built in accordance to the simulated electrostatic lens system. The electrodes were produced out of stainless steel, for insulating parts Macor (a glass-ceramic well suited for UHV conditions) was used. To avoid surface charges on the insulators, which could severely influence the electric field, the electrode geometry was designed in a way that there is no direct connection from any point of the ion beam to the insulator surfaces (c.f. Fig. \ref{fig:lens_crosssection}).
In addition, the minimal distance between the electrodes has been limited to \SI{1}{mm} in vacuum and \SI{7.25}{mm} at Macor insulators. This ensures a sufficient dielectric strength of about \SI{6}{kV} \cite{Drummond.1984}. In contrast to all the other electrodes, the positive extractor consists out of transparent glass, coated with \SI{80}{nm} of indium tin oxide (ITO). By using such a glass electrode, we gain optical access to the MOT parallel to the optical axis of the ion optics while being able to apply a voltage.

The ion detection with high temporal and spatial resolution is done by a combination of a microchannel plate (MCP) and a delay line detector (DLD)\cite{Sobottka.1988, Friedman.1996}. The electrical signals of the DLD are discriminated by a set of constant fraction discriminators and recorded with a time to digital converter (TDC). An ion-optical image is typically achieved by accumulating the ion events at the detector for a given amount of time and plotting a histogram of the detected ion positions. The detection efficiency of the system is expected to be mainly given by the MCP, which typically lies in the range of 60-85\% for positive ions with the kinetic energy used in our setup \cite{LadislasWiza.1979}.

\section{\label{sec:meas}Measurement}
In order to prove the ability of our ion optics to image cold atoms, defined test structures have been applied. Therefore, we spatially structured the ionization laser since the photoionization rate is proportional to the laser intensity.

We imaged a resolution test target (USAF1951) illuminated by the ionization laser onto the MOT with an achromatic lens positioned outside the vacuum chamber. The magnification of this light-optical imaging system was $M_L=1.14$, the optical axis of the imaging is parallel to the optical axis of the ion optics and goes through the ITO coated glass electrode. The test target consists of several groups of elements with varying size. Each element consists of three vertical and three horizontal bars with a distance identical to their width. Obviously, our very simple light-optical projection has a limited quality, so the smallest usable structures has a center-to-center distance of \SI{20}{\micro\meter} at the position of the MOT. Hence, this method is suitable for testing the ion imaging at small magnifications with relatively large structures.

\begin{figure}
	\includegraphics{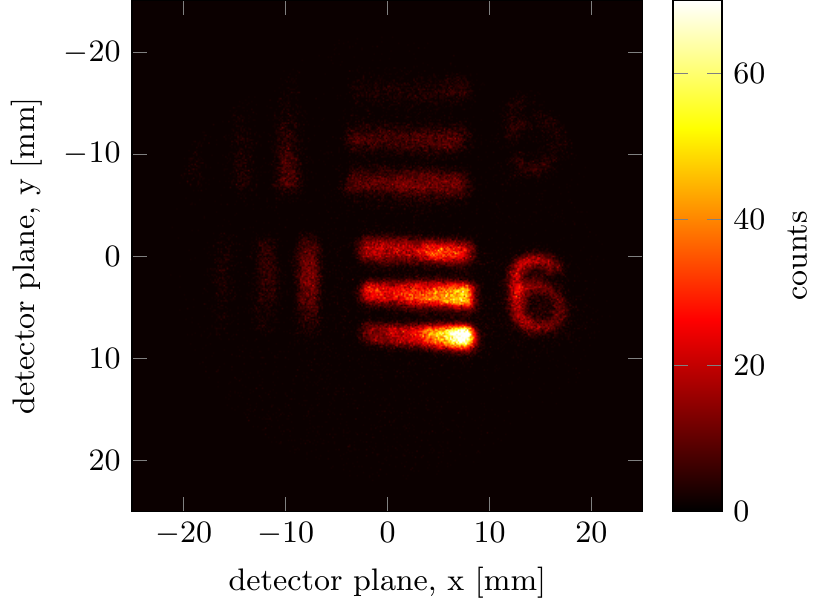}
	\caption{\label{fig:USAF_meas_single}Ion image of a USAF1951 test structure that was projected with the ionization laser onto the MOT. The visible bar structure (number six) has a center to center distance of \SI{80}{\micro\meter} at the position of the MOT. The voltages at the center electrodes of the einzel lenses were $U_{\text{L3}}=\SI{300}{V}$, $ U_{\text{L1,2,4}}=\SI{-2.4}{kV}$. The magnification is about 50.}
\end{figure}

\begin{figure}
	\includegraphics{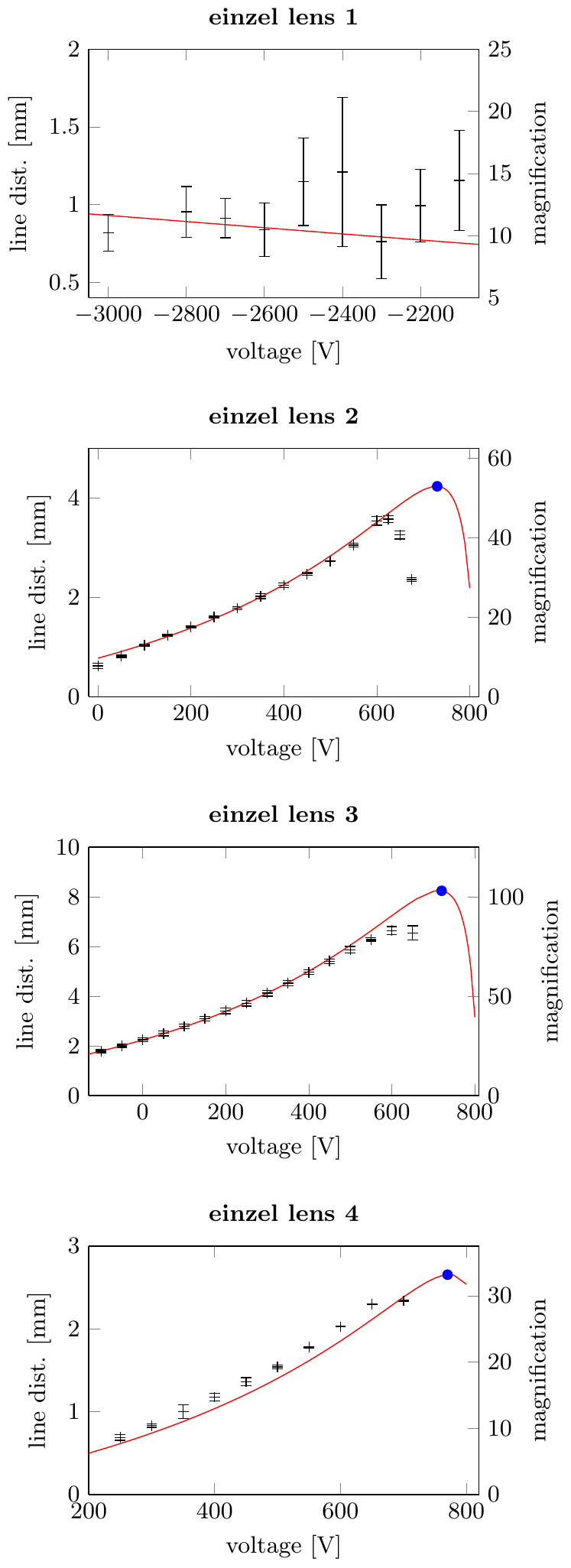}
		\caption{\label{fig:USAF_meas_all}Center to center distance of the test structure from Fig. \ref{fig:USAF_meas_single} on the detector in dependence on the voltage applied to the center electrode of the corresponding einzel lens. The measurement was evaluated by determining the distances between multiple line pairs of the corresponding structure and is shown in black with error bars. The results of the simulation are depicted as solid red lines. The voltage setting for the sharpest image (minimum of the point spread function) is marked with a blue dot (if any). For other settings, the voltage of at least one other lens has to be adapted in order to achieve the sharpest possible image. }
\end{figure}

To test the imaging of each einzel lens individually, we measured the ion image of a test structure with only one active einzel lens, all the other einzel lenses were ``switched off" by applying the drift-tube voltage of \SI{-2.4}{kV} to the center electrode. At the extraction electrodes, voltages of $\pm\SI{500}{V}$ were applied. Because of that, the first einzel lens still has a focussing effect even with the center electrode at \SI{-2.4}{kV}. From the USAF1951 target, we used the structure 6 of group 3, that has a center to center distance of $M_L \cdot \SI{70.16}{\micro\meter} \approx \SI{80}{\micro\meter}$. The corresponding image, as measured with our ion microscope at a magnification of 50, is shown in Fig. \ref{fig:USAF_meas_single}.
It nicely visualizes the original target pattern and yields a distortion free and sharp image. We now imaged this structure for different voltages at the center electrode of the active einzel lens and measured the line distance in the detector image. The results are shown in Fig. \ref{fig:USAF_meas_all} and unveil the voltage dependent magnifications of each einzel lens.
All measurements (black) show good agreement to the values extracted from the simulations (red). Only at high voltages, the measurements deviate from the simulations. Partly, this can be explained by inaccuracies in the mechanical assembly, asymmetries and voltage deviations, but it is also likely, that the extraction field in the experiment differs from the simulations. The simulations only consider a rotational-symmetric electrode geometry but in reality, there are also other, non-rotational-symmetric parts that can influence the electric field (for example the MOT-coil holders). Another source for the deviation can be the position of the MOT between the extractor electrodes. In the simulations, the ions start exactly in the center between the extractors, but in the experiment the position of the MOT relative to the extractors cannot be determined exactly.

\begin{figure}
	\includegraphics{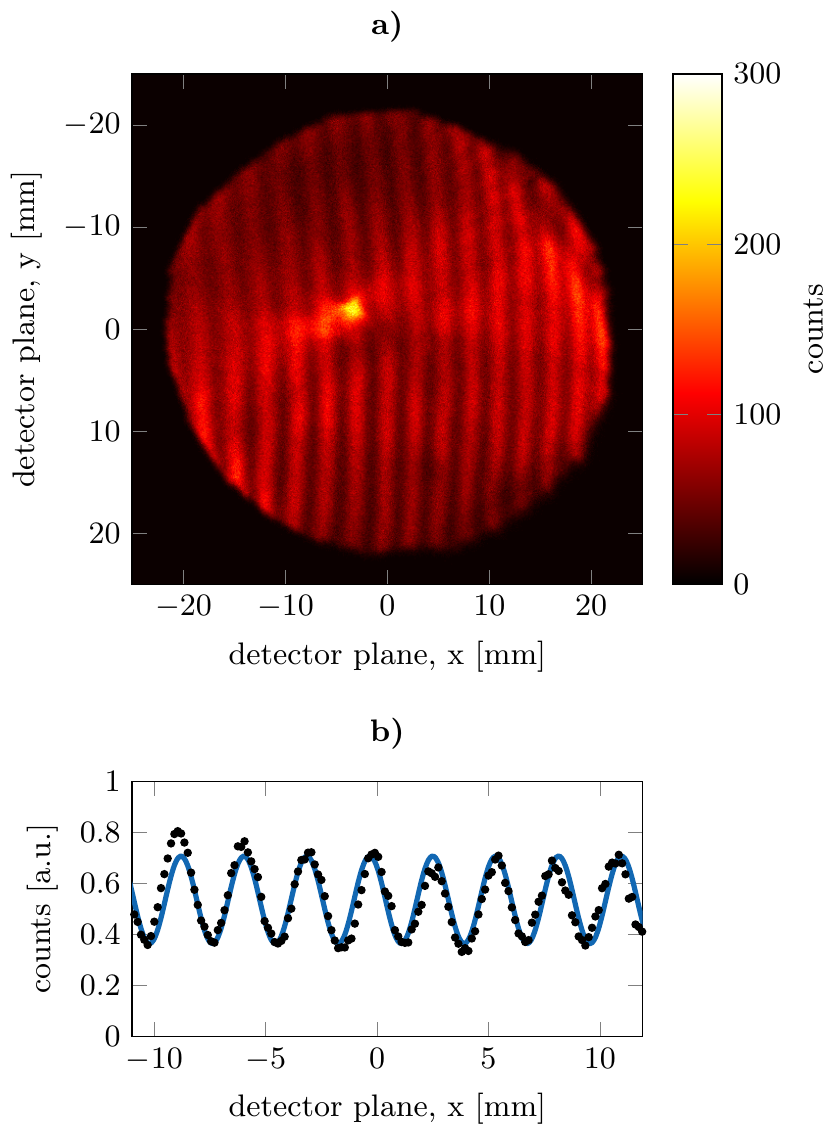}
	\caption{\label{fig:fringes2_7um}a) Ion-optical imaging of a MOT onto the MCP with the ionization laser intensity spatially modulated by an interference structure with fringe spacing of \SI{2.7}{\micro\meter}. The magnification is 1050. The bright spot in the middle is produced by ions that are generated outside of the detection region and whose trajectories get bent back into the image. b) Line scan of the measurement summed up over $y\in \left[ \SI{11.6}{mm}, \SI{13.1}{mm}\right]$. The blue line is a a fit with a sine-function plus offset. The contrast of the fringes (calculated out of the fit) is 0.32.}
\end{figure}

The USAF1951 target is well suited to characterize the relatively small magnification of a single einzel lens but cannot be used to characterize high magnifications and the resolution limit of the ion optics. For smaller, but still well defined structures, an optical lattice as described in \cite{Li.2008} was implemented by superimposing two parallel beams of the ionization laser with a lens onto the MOT. By varying the distance between the parallel beams, the angle between the beams behind the lens and with that the lattice spacing of the emerging interference pattern can be changed. As the beams go through the glass electrode, the maximum beam distance and with that the minimal lattice spacing is limited to about $\SI{2.7}{\micro\meter}$. With this setup, the predicted magnification of 1000 could be achieved and structures down to the lattice spacing limit could be imaged easily as displayed in Fig. \ref{fig:fringes2_7um}a. 
There is a small curvature of the fringes visible, that can be explained by monochromatic aberrations (see Fig. \ref{fig:sim_spherical_astigm}). The fringes are bent in the same direction over the whole image plane and furthermore there is a slight increase in magnification from one side of the detector region to the other, resulting in a changing fringe separation. Both effects indicate, that one or more electrodes are not positioned exactly rotationally symmetric to the optical axis. A line scan of the measurement is depicted in Fig. \ref{fig:fringes2_7um}b, showing a fringe contrast of 0.32. However, it is difficult to extrapolate a resolution limit out of this contrast, due to the large depth of field of our imaging system. Ions from areas where there is little to no overlap of the beams producing the lattice structure get also imaged onto the detector plane, thus adding an offset to the ion signal. Furthermore, the trajectories of ions produced outside the detection region can get bent back onto the detector. For the future, a reduction of the ionization area to a plane - for example with a light sheet - is advisable. Nevertheless, the \SI{2.7}{\micro\meter} lattice is clearly resolvable, which implies that the actual resolution limit lies lower.

Testing the ultimate resolution limit of the system would require smaller structures to be imprinted onto the atoms. This could be done by using the standing wave pattern of a retro-reflected ionization laser. In our setup, that would result in a lattice structure with \SI{240}{nm} fringe spacing. The structure would probably have to be mechanically stabilized to the ion-optical system to minimize the effect of vibrations. Another crucial point is the stability of the voltage sources of the electrodes, especially for ions starting far away from the optical axis. Simulations show that a stability of $\pm 0.05\%$ has to be maintained in order to not limit the resolution of the system. In addition to that, deviations from rotational symmetry can also limit the resolution as well as the characteristics of the delay line signals and detector electronics. As the lattice spacing of the retro-reflective configuration is more than one order of magnitude smaller than the smallest structure demonstrated here, a more promising approach would be to start with structure sizes closer to the demonstrated \SI{2.7}{\micro\meter}. This could be done with a variable standing wave similar to the one we used but with a higher maximum angle between the laser beams, which was not possible in our setup due to limited optical access. Another approach would be to use a different photoionization path involving a wavelength suitable for the desired lattice spacing.

\section{\label{sec:conclusion}Conclusion}
In summary, we have presented an ion-optical setup to image and magnify cold atom clouds. The system consists out of four electrostatic lenses and a MCP for ion detection. Simulations show, that the resolution should be better than \SI{100}{nm} with a maximum magnification of 1000. Therefore, it surpasses standard light optical imaging techniques. The ion optics was realized experimentally and integrated into a cold atom setup. It was characterized by photoionization of atoms out of a magneto-optical trap. The results are in good agreement to the simulations and show, that the system can be used to study ultracold atom clouds.

Although, the ion microscope was demonstrated here on a continuously operated MOT, it can be easily applied to investigate ultracold quantum gases and Bose-Einstein condensates. Then, the microscope can be operated in a quasi non-destructive mode by ionizing only a small subset of particles. If this subset resembles the whole quantum gas, global properties like atom number, density and temperature can be extracted from the ion microscope. With the microscope being able to operate continuously, also the observation of dynamical processes in quantum gases, like vortices, solitons, collective excitations or oscillations come into direct reach. Furthermore, the system should be well suited to investigate local statistics in quantum gases via temporal and spatial correlation analysis. Such correlation measurements become even more important in the context of Rydberg gases, where many-body effects like the Rydberg blockade \cite{Singer.2004,Tong.2004}, facilitation \cite{Lesanovsky.2014}, crystallization \cite{Schauss.2015} or molecule formation \cite{Bendkowsky.2009} could be directly observed. The cold atom microscope is thus perfectly suited to investigate physics beyond the standard mean-field approach.

\appendix*

\begin{acknowledgments}
We gratefully acknowledge financial support from Deutsche Forschungsgemeinschaft through SFB TRR21, from Baden-W{\"u}rttemberg Stiftung through 'Kompetenznetz Funktionelle Nanostrukturen' and from FET-Open
Xtrack Project HAIRS. M.S. acknowledges financial support from Landesgraduiertenf\"{o}rderung Baden-W\"{u}rttemberg. The authors would also like to thank C. Zimmermann and P. Federsel for helpful discussions and C. Billington for assistance during the set-up of the experiment.
\end{acknowledgments}

\bibliography{literature}

\end{document}